\documentclass[a4paper,12pt]{article}
\usepackage{amsmath,amssymb,amsfonts}
\usepackage[dvips]{graphicx}
\usepackage{tabularx}
\usepackage{bbold}
\usepackage{setspace}
\usepackage{cite}

\title{Causal Effect Estimation Methods}

\author{Priyantha Wijayatunga\footnote{The author gratefully acknowledges the financial support of the Swedish Research Council through the Swedish Initiative for Microdata Research in the Medical and Social Sciences (SIMSAM)  and Swedish Research Council for Health, Working Life and Welfare (FORTE).} \\  Department of Statistics  Ume\r{a} School of Business and Economics\\ Ume\r{a} University Ume\r{a} 90187 Sweden \\\emph{Email: priyantha.wijayatunga@stat.umu.se}} 

\begin{document}
\maketitle

\begin{abstract} 
Relationship between two popular modeling frameworks of causal inference from observational data, namely, causal graphical model and potential outcome causal model is discussed. How some popular causal effect estimators found in applications of the potential outcome causal model, such as inverse probability of treatment weighted estimator and doubly robust estimator can be obtained by using the causal graphical model is shown.   We confine to the simple case of binary outcome and treatment variables with discrete confounders and it is shown how to generalize results to cases of continuous variables.  
\end{abstract}

\noindent%

\newtheorem{assump}{Assumption}
\newtheorem{prop}{Proposition}
\newtheorem{lem}{Lemma}
\newtheorem{thrm}{Theorem}
\newtheorem{cor}{Corollary}
\newtheorem{exam}{Example}
\newtheorem{mydef}{Definition}

\section{Introduction}

In many real world situations it is of interest the estimation of causal effect of some treatment on a certain outcome. The causal effect of taking a certain medicine for a certain disease by the patients and that of participation in a certain job training program by unemployed individuals in order to find employment in the future are two examples among many in medical and socio-economic contexts respectively, among many others in a lot of disciplines. Sometimes it may be unethical or infeasible to assign each subject either to the treatment or to the control randomly in order to perform a randomized study that is considered as the gold standard to estimate the causal effect of the treatment. However it may be of interest of socio-economic policy makers, medical professional, etc., to evaluate the causal effect of their treatments of interest in order to plan for the future. In the absence of the randomized assignment of the treatments they may only have observed data on collection of subjects  who either have taken the treatment or not.  When  the effect of a treatment on an outcome needs to be identified from such observational data sample it needs to control for (condition on) the confounders, i.e., subgroups with the same confounder values in the treatment group and those in the control group should be compared through their empirical mean values of the outcome and then it should be  taken the weighted average of them where weights are observed proportions of sizes of the subgroups in the data sample to evaluate the average causal effect of the treatment.  For simplicity assume all confounders are discrete. Note that the confounders are factors that affect the subjects to take the treatment or not while simultaneously affecting the subjects' outcome in some way, therefore the effect of the treatment is confounded with the effects of these confounding factors when they are present. So, these unnecessary effects should be removed otherwise the estimate of the average treatment effect is biased.  One can see that here the implicit assumption is that within each subgroup of confounder value the  treatment assignments are assumed to be randomized, therefore comparisons are done subgroup-wise. But this assumption is true when a 'sufficient' set of confounders, perhaps not all of them are considered.  

However sometimes controlling for the confounders can be difficult, for example, if they are high dimensional then it may be difficult to find treatment and control subgroups of subjects of sufficient sizes with same confounder values.   A popular way to increase the sizes of these treatment and control subgroups that should be compared is to  use so-called propensity scores \cite{RR1983}. The propensity score is the  conditional probability of receiving the treatment given the values of observed pre-treatment confounding covariates of the treatment and the outcome. Among others, they are used in the causal inference method  of potential outcome framework \cite{RD1974,HP1986} for matching subgroups of treated subjects with those with untreated, usually called stratification of data sample,  for estimating the causal effects of the treatments.  

Finding a 'sufficient' set of  condounders on which the comparison should be done  is somewhat problematic and the potential outcome framework offers no clear way to do it even when all pretreatment confounders of the treatment and the outcome are available. Note that one does not need to control for all the confounders since when some of the them are considered then some of the others may become redundant. However causal graphical modeling framework of Pearl and his colleagues (see \cite{PJ2009} and references therein) offers one called 'back door criterion' to choose a set of sufficient covariates in order to identify the causal effect, i.e., to estimate without bias.  When a graphical model is done on the treatment variable and outcome variable and all their assumed causal factors, both direct and indirect, the criterion can find a sufficient set of covariates on which one should control for estimation of the causal effect. Then such a set is called 'admissible' or 'deconfounding' set. However the selected set is only sufficient for all the causal factors  that are assumed but may not be sufficient if some causal factors of treatment and outcome are omitted.  And considering some covariates as confounders by ignoring such criterion or similar one can cause introduction of further bias (p. 351 of \cite{PJ2009}).   So, in our analysis we confine to the case of that taken confounders make a superset of an admissible set and stated otherwise. 

Often these two camps of causal inference methods have a lot of disagreements between them, especially the applied users of them. However developers of the two frameworks, if not theoreticians in them have remarked the relationship between them. One such instance is reported in the journal "NeuroImage" under the section "Comments and Controversies"  about applying two modeling frameworks for brain image data \cite{RSC2011, LS2011, PJ2011, LS2013, GC2013}. Therein Pearl argues that his group (in his words) has proved that two frameworks are logically equivalent in the sense that a theorem in one is a theorem in the other and an assumption in one has a parallel interpretation in the other. And Glymour argues that (in his words) the potential outcome model is an special case of the causal graphical model but with twists that make causal estimation impossible except in restricted contexts. And others in the debate are of the opinion that the two frameworks are close to each other. Though such arguements are around among the theoreticians of the two frameworks, the applied users still seem to be unconvinced about it, and therefore they treat that they are very different frameworks and often one is supeiror than the other. Or even worst, one gives wrong answers while  only the other gives correct answers. It is rare that both frameworks are applied for same data. Furthermore due to different numerical estimation methods one may obtain two numerically different causal effect estimates when the two frameworks are used.   

Here we show that two frameworks are equivalent in most contexts in the sense that both give same analytical expressions for causal effect estimates or rather any causal effect estimate in one modeling framework can be obtained from the other. Since causal effect estimates are dependent on estimated  probabilities because  they are functions of statistical conditional expectations of outcome variable there can be differences in causal effect estimates numerically if the used probabilities are estimated differently. But there are reasons, at least operationally, to favor the graphical modeling framework over the other, for example, it can be computationally efficient, for example, through controlling for a sufficient set of confounders rather than doing so for all the assumed confounders. We show their equivalence at the basic level of their application. Furthermore since the potential outcomes model has many forms causal effect estimators we show how they can be derived through the graphical modeling framework, thus providing some insight into the estimators.  So, our discussion here can be useful not only for researchers  in these two modeling frameworks but also especially for the users of them to understand each other.   

\section{Observational Studies}

We consider the simple situation where one is interested in evaluating the effect of some exposure or treatment on a certain outcome that can either be a success or a failure. Let us denote the  treatment by a binary variable $Z$ where $Z=1$ when the treatment is implemented and $Z=0$ when it is not and the outcome by a binary variable $Y$ where $Y=1$ when a success is observed and $Y=0$ when a failure is observed for each subject concerned. In the potential outcome framework for causal inference  it is accepted existence of pair of  potential outcome variables, say, $(Y_1, Y_0)$ where $Y_i$ is the outcome that would have been observed  had the treatment $Z=i$ for $i=1,0$. Note that then the observable outcome $Y$ satisfies  the relation $Y=ZY_1 + (1-Z)Y_0$. Then a randomized experiment is when the potential outcomes are independent of treatment assignment, written as $(Y_0,Y_1) \perp Z $; each subject receives treatment without considering its future outcome. Then average causal effect for the population $\tau$ is defined as follows. 
\begin{eqnarray*}
\tau & = & E[Y_1] -  E[Y_0] \\ 
           &=&   E[Y_1 \vert Z=1] -  E[Y_0 \vert Z=0] \textrm{  since }  (Y_0,Y_1) \perp Z \\
           &= &  E[Y \vert Z=1] -  E[Y \vert Z=0] 
\end{eqnarray*}
Here we assume that  $0<P(Z=1)<1$, i.e.,  in our sample of data we have both treated and untreated subjects. If it is not the case then we are not able to estimate $\tau$ since then only one of quantities in the expression is known.

But in observational data the independence assumption $(Y_1,Y_0) \perp Z$ may not hold because subjects do not receive the treatment independent of their future outcomes, therefore characteristics of subjects in the treatment group may differ from those of the control group. This is a situation where the treatment effect is confounded with some external factors, i.e.,  the treatment and the outcome are confounded. Therefore the treatment group and control group cannot be compared directly to evaluate the effect of the treatment. Then the assumption is modified and it  says that the potential outcomes are conditionally independent of the treatment assignment given some confounding factors that makes (a superset of) an admissible set for confounding. When this set of confounders are denoted by multivariate variable $X$  then the assumption is written as $(Y_1,Y_0) \perp Z \vert X$ and it is sometimes called the assumption of no unmeasured confounders to mean that all the confounding effects are removed by $X$. In addition, for inference, similar to randomized experiment it needs to have $0<P(Z=1 \vert X) <1$, which is called assumption of common support.  That is, for each configuration (stratum) of $X$, we should have both treated and untreated subjects otherwise, say for example, if $P(Z=1 \vert X=x_1)=1$ in our data sample then the causal effect for the subgroup with $X=x_1$ may not be calculated.  Recall that we assume that $X$ is discrete, therefore any continuous covariate is discretized.  That is, in each stratum of $X$ the treatment assignments are as if they are  randomized and we have data on both treated and untreated subjects. This is to say that  in observational data our objective is to mimic the randomization within each stratum of $X$.  Therefore, firstly one should find a sufficient set of confounders $X$. However this assumption cannot be tested even if all the potential confounders are found.  

Now let us define that individual causal effect for an individual, say, $j$ with $X=x$ is $\tau^{j}(x)=Y_1^j -Y_0^j$. The $j^{th}$ individual is the $j^{th}$ data case of the sample and throughout any quantity referring to it is denoted with the superscript $j$ attached to the respective quantity. But it is clear that no subject has both the values of $Y_1$ and $Y_0$ observed therefore we cannot have $\tau^j(x)$ numerically. So we need a mechanism to get it but it is right at our hands; the randomization of the treatment assignments within  each stratum of $X$, the assumption of no unmeasured confounders (this is also called the assumption of strong ignorable treatment assignment \cite{RR1983}). That is, within any stratum $X=x$ if we know a subject is treated ($Z=1$) we observe $Y_1=Y$ but $Y_0$ is not known, but the latter can be known by any other subject in the stratum who is not treated ($Z=0$); two quantities are conditionally exchangeable.  Here the word 'conditionally' is to mean that within the stratum.  And similarly for any subject that is not treated ($Z=0$). Therefore, as if the observed data are from randomizations within each level $x$ of $X$, we can calculate the  average causal effect for the subpopulation of all individuals with $X=x$, say, $\tau(x)$ by 
\begin{eqnarray*}
\tau(x) & = & E[Y_1 \vert X=x] -  E[Y_0 \vert X=x] \\ 
           &=&   E[Y_1 \vert X=x,Z=1] -  E[Y_0 \vert X=x,Z=0]  \textrm{ since } (Y_1,Y_2) \perp Z \vert X \\
           &= &  E[Y \vert X=x,Z=1] -  E[Y \vert X=x,Z=0] 
\end{eqnarray*}
where the expectation $E$ should be taken over whole subpopulation with $X=x$. Since this mechanism applies for all the strata of $X$, we can calculate  the average causal effect for the whole population, say, $\tau$  
\begin{eqnarray*}
\tau &=& E_x [ E[Y \vert Z=1, X=x] -  E[Y \vert Z=0, X=x]] \\
      &=& \sum_x  \sum_y y p(Y=y \vert X=x,Z=1)p(X=x) \\
      & & - \sum_x  \sum_y y p(Y=y \vert X=x,Z=0)p(X=x)
\end{eqnarray*}
It is sufficient to estimate accurately the probabilities $p(Y=y \vert X=x,Z=z)$ and $p(X=x)$ for $Z=0,1$ and for all values of $ X$ in order to estimate $\tau$ accurately but due to its definition it is not necessity. For example, if some forms of errors have been introduced in calculation of $p(Y=1 \vert X=x, Z=1)$ then similar errors in calculation of $p(Y=1 \vert X=x, Z=0)$  may make sure that the correct value for $\tau$ is obtained. For these types of reasons or similar ones sometimes researchers claim that even models, for example, those for conditional probabilities, are misspecified correct estimates for causal effects can be obtained. But here we avoid discussion on this topic.

The above estimate for $\tau$ is analytically equal to that we get by the estimation of the causal effects using interventions in causal graphical models (also called do-calculus) \cite{PJ2009,LR2002}, that is another popular framework for the task,  therefore two frameworks are equivalent in this case. To recall the reader with this calculus, first define the distribution with conditioning by intervention or action; if we have observed a random sample of data on a set of variable, say, $X_1,...,X_n$, we can find the probability distribution of the set of variables, say, $p(x_1,...,x_n)$. We can have a factorization of probability distribution $p(x_1,...,x_n)$; let it be that  $p(x_1,...,x_n)= \prod_{i}^{n} p(x_i \vert pa_i)$ where $pa_i \subseteq \{x_1,...,x_{i-1}$\} with the exception of $pa_1= \Phi$ (empty set) using some conditional independence assumptions within $X_1,...,X_n$. Note that to have a causal representation in the factorization one can use, for example, the time order to index the variables such that cause variables have higher indices than those of effect variables'. Then, for $i=1,...,n$ the probability distribution of $\{X_1,...,X_n\} \backslash \{X_i\}$ when $X_i$  is intervened to a particular value of it, say, $x_i$, written as $do(X_i=x_i)$, denoted by $p(\{x_1,...,x_n \} \backslash \{x_i\} \vert do(X_i=x_i))$ is defined as follows; 

\begin{eqnarray*}
p(\{x_1,...,x_n\} \backslash \{x_i\} \vert do(X_i=x_i)) &=& \frac{p(x_1,...,x_n)} {p(x_i \vert pa_i)}=\prod_{k=1:k \neq i}^{n} p(x_k \vert pa_k) \\
 & \neq & \frac{p(x_1,...,x_n)} {p(x_i )} =\frac{1}{p(x_i)}\prod_{k=1}^{n} p(x_k \vert pa_k)  \\
&=& p(\{x_1,...,x_n\} \backslash \{x_i\} \vert X_i=x_i) 
\end{eqnarray*}
where the last expression is corresponding conditional probability distribution when we have observed $X_i=x_i$,  which is generally different from that of conditioning by intervention.  

\small
\begin{figure} 
\begin{center}
\setlength{\unitlength}{1mm}
\begin{picture}(100,40)(-50,-10)
\thicklines
\put(0,20){\circle{7}}
\put(-2,19){$X$}

\put(-20,0){\circle{7}}
\put(-22,-1){$Z$}

\put(20,0){\circle{7}}
\put(18,-1){$Y$}

 \put(-2,17){\vector(-1,-1){15}}
 \put(2,17){\vector(1,-1){15}}
 \put(-16.5,0){\vector(1,0){33}}

\put(-22, -10){$p(y,z,x)=p(x)p(z\vert x)p(y\vert x,z)$}
\end{picture}
\end{center}
\caption{ \label{simple.bn}  Bayesian network for causal model}
\end{figure}
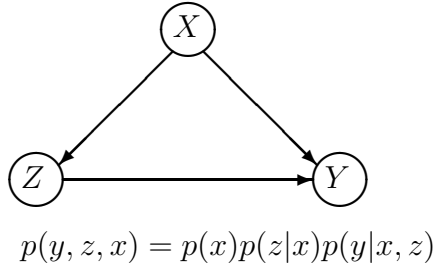
\normalsize 
The causal relationships between $X$, $Y$ and $Z$ in our context can be represented as a causal network model $p(y,z,x)=p(x)p(z\vert x)p(y\vert x,z)$ as shown in the Figure \ref{simple.bn}. And if we intervene on $Z$ as $do(Z=z)$ for $z=0,1$, then the intervention distribution  
\begin{eqnarray*}
p(Y=y, X=x \vert do(Z=z) &=& \frac{p(X=x)p(Z=z \vert X=x)p(Y=y \vert Z=z,X=x)}{p(Z=z \vert X=x)} 
\end{eqnarray*}
So we have $p(Y=y \vert do(Z=z)) = \sum_x p(Y=y \vert Z=z,X=x)p(X=x)$. The causal effect of the treatment option $Z=1$ compared to the control option $Z=0$ is defined as 
\begin{eqnarray*}
\rho &= & \sum_y y p(Y=y \vert do(Z=1)) - \sum_y y p(Y=y \vert  do(Z=0)) \\
       &=&  \sum_y y \sum_x p(Y=y \vert Z=1,X=x)p(X=x) - \sum_y y \sum_x p(Y=y \vert Z=1,X=x)p(X=x)  \\
      &=&  \tau  
\end{eqnarray*}
So we have seen that strong ignorable treatment assignment assumption  in potential outcome model is equivalent to implementing intervention operations in probability distributions when  the confounding factors are the same in both cases, i.e., they yield analytically the same causal effect estimates. In fact, for the above one can see that the probability distribution of the potential outcome of a hypothetical treatment assignment under the strong ignorability assumption and that of  the outcome of the intervention of same value are the same. For $i,j=0,1$,
\begin{eqnarray*}
p(Y_i=y)&=& \sum_x p(Y_i=y \vert x)p(x) = \sum_x p(Y_i=y \vert Z=j,x)p(x) \\
             &=& \sum_x p(Y=y \vert Z=i,x)p(x) =p(Y=y \vert do(Z=i))
\end{eqnarray*}

Now suppose the case where treatment has also an indirect effect on the outcome in addition to its direct effect. Suppose effect of $Z$ on $Y$ is also mediated through $Z'$ and a  set of confounders among causal relationships between them is denoted by $X$ such that $X$ is the union of distinct sets of confounders $X_1$, $X_2$, $X_3$ and $X_4$ where $X_1$ and $X_4$ are the set of all direct confounders for direct causal relation between $Z$ and $Z'$, $Z$, $X_2$ and $X_4$ are those between direct causal relation between $Z'$ and $Y$, and $X_3$ and $X_4$ together complete the set of all confounders for the  indirect causal relation between $Z$ and $Y$.  Here we have taken all the confounders rather than respective admissible sets for simplicity. Let us define the potential outcome $Y_{ij}$ the outcome that would have observed had $Z=i$ and $Z'=j$  and then $Y_i=Z' Y_{i1}+(1-Z')Y_{i0}$ for $i,j=0,1$ and $Y=Z Y_1+(1-Z)Y_0$. Then we have strong ignorability assumptions  $Z_1',Z_0' \perp Z \vert \{X_1,X_4\}$ and $Y_{i1},Y_{i0} \perp Z' \vert \{Z=i,X_2,X_4 \} $ for $i=0,1$ for the direct causal relationships between $ Z \rightarrow Z'$ and $Z' \rightarrow Y$ respectively. But they do not imply ignorability assumption for $(Y_1,Y_0)$ and $Z$. So we need to assume, for example, safely  that $Y_1,Y_0  \perp Z \vert X $. Note that there is no obvious way to take a subset of $X$ as the conditioning set. In this case also we get, for $i=0,1$, $p(Y_i=y) =  \sum_{x} p(Y=y \vert Z=i,x)p(x) $ for $i=0,1$. And in the causal graphical model
\begin{eqnarray*}
p(x_1,...,x_4,z,z',y)   &=& p(x_1,...,x_4)p(z \vert x_1,x_3,x_4)p(z' \vert z,x_1,x_2,x_4)p(y \vert z',z,x_2,x_3,x_4) \\
p(x_1,...,x_4,z',y \vert do(z))   &=&p(x_1,...,x_4)p(z' \vert z,x_1,x_2,x_4)p(y \vert z',z,x_1,x_2,x_3,x_4) \\
p(y \vert do(z)) &=& \sum_{x_1,..,x_4,z'} p(x_1,...,x_4)p(z' \vert z,x_1,x_2,x_4)p(y \vert z',z,x_1,x_2,x_3,x_4) \\
                       &=&   \sum_{x_1,x_2,x_3,x_4} p(x_1,x_2,x_3,x_4)p(y \vert z,x_1,x_2,x_3,x_4) 
\end{eqnarray*}
Therefore,  $p(Y_i=y) = p(y \vert do(Z=i)) $ for $i=0,1$.   So we have seen the two frameworks are yielding same causal effect estimates, therefore two frameworks are equivalent in this case too. However,  since $p(y \vert do(z))=\sum_{x_2,x_3,x_4} p(x_2,x_3,x_4)p(y \vert z,x_2,x_3,x_4)$ using graphical model is more efficient compared to doing so the potential outcome framework.

Since one can encounter situations where the causal structures of the phenomena are complex, it is advisable to use the causal graph interventions for  estimation of desired causal effect. If the confounding factors taken into consideration in the potential outcome model and the graphical model are the same then both models yield analytically the same causal effect estimates. 

\section{Some Differences in Two Modeling Frameworks} \label{SomDiff}

As seen earlier, in order to have same numerical causal effect estimates in both frameworks they should include supersets of similar admissible sets of confounders and  same probability density estimates. However, researchers who use the potential outcome model tend to include pretreatment covariates that are associative but not causal with both $Z$ and $Y$ too as confounders. This can induce spurious bias as shown in literature using the graphical modeling framework. Such factors may not be direct confounders but they are said to be inducing so-called M-bias in casual effect estimation. Therefore researchers argue that they should be neglected in causal effect estimation \cite{SI2009, RD2009, SA2009, PJ2009a}. However when a pretreatment covariate that is associative with both treatment and outcome is found this may indicate that either there is another unmeasured confounder or two dependent unmeasured confounders in the system, not necessarily two independent confounders as considered in the above debate. If former two  are the cases (either single unmeasured confounder or two dependent confounders) whether conditioned on associative confounder or not causal effect estimates are biased. In a forthcoming paper  \cite{WL2014} it is shown that in these two cases it is more beneficial to condition on the associative confounder than not doing so. We avoid discussion on this topic here. 

Another difference is caused by discriminative and generative estimation of probabilities  where in the potential outcome model often individual conditional probabilities are estimated discriminatively, for example, using logistic regression for propensity score estimation whereas in the graphical model often joint likelihood is maximized to obtain component conditional probabilities of the factorization of joint density of $Z,X$ and $Y$. The factorization $p(X=x,Z=z,Y=y)=p(X=x)p(Z=z \vert X=x)p(Y \vert Z=z, X=x)$ includes propensity scores and therefore if two estimation methods yield two numerically different estimates for propensity scores then it can result in two different causal effect estimates.  See below for further comments.

\section{Some Causal Effect Estimators} 

Let us see how the graphical model estimator can be used to derive the causal effect estimators such as   inverse probability of treatment weighted estimator,  stratified estimator and doubly robust estimator commonly found in   the potential outcome model applications. In the following we avoid direct definition of those estimators but derive them by manipulation of the graphical model estimator.  

\subsection{Inverse Probability of Treatment Weighted Estimator}

The graphical model causal effect estimator $\rho$ is equivalent to inverse probability of treatment weighted estimator (\emph{IPTW}) described \cite{RH2000}. 
\begin{eqnarray*}
\rho &= & \sum_y y p(Y=y \vert do(Z=1)) - \sum_y y p(Y=y \vert  do(Z=0)) \\
      &=& \sum_y y \sum_x p(Y=y \vert Z=1,X=x)p(X=x)  \\
      & &- \sum_y y \sum_x p(Y=y \vert Z=0,X=x)p(X=x) \\
      &=& \sum_y y \sum_x \frac{p(Y=y, Z=1,X=x)}{p(Z=1 \vert X=x)} - \sum_y y \sum_x \frac{p(Y=y,Z=0,X=x)}{p(Z=0 \vert X=x)} \\
     &=&  \sum_x \frac{p(Y=1, Z=1,X=x)}{p(Z=1 \vert X=x)} - \sum_x \frac{p(Y=1,Z=0,X=x)}{p(Z=0 \vert X=x)} \\
     &=&  \sum_x \frac{1}{e(x)}\frac{N(Y=1, Z=1,X=x)}{N}  -\sum_x \frac{1}{1-e(x)}\frac{N(Y=1, Z=0,X=x)}{N} \\
    &=&  \sum_i \frac{1}{e(x^i)}\frac{I(Y^i=1) I(Z^i=1) I(X^i=x^i)}{N}  -\sum_x \frac{1}{1-e(x^i)}\frac{I(Y^i=1) I(Z^i=0) I(X^i=x^i)}{N} \\
    &=& \frac{1}{N} \sum_i \frac{Z^i Y^i}{e^i} -\frac{1}{N} \sum_i \frac{(1-Z^i) Y^i}{1-e^i} = IPTW
\end{eqnarray*}
where $N(.)$ denotes the number of data cases satisfying its arguments, $I(.)=1$ when its argument is true and $I(.)=0$ otherwise and $p(Z^i=1 \vert X^i=x^i)=e(x^i)=e^i$.  Therefore, analytically the graphical model  intervention estimator is the \emph{IPTW} estimator. However often they can be different numerically, for example, when the propensity score estimates, $e(x)$ for all $x$, differ in the two contexts as discussed in Section \ref{SomDiff}.

\subsection{Stratified Estimator}

Essentially we can obtain the propensity score stratified estimator \cite{WM2012} from \emph{IPTW} since it is just a stratification of range of propensity score values into several bins where within each bin it is assumed the propensity scores are approximately the same. In fact it is an algebraic simplicity (summing up fractions by assuming some of them have equal denominators) but what it is important to note is that  in the stratified estimator those common propensity scores for corresponding sets of approximately equal propensity scores are estimated by the sample proportions of treated subjects related to  those propensity  scores. In fact, the estimates are maximum likelihood estimates for $P(Z \vert X')$ from the likelihood for the joint density where $X'$ obtained from $X$ through a 'new' definition on the state space of $X$. For clarity we can see how stratified estimator is related with \emph{IPTW} estimator. Note that since it is emplicit that common propensity score values are in fact used in stratified estimator, for most applied researchers it is not clear about it. Suppose we write propensity score estimates in increasing order for all the subjects, say, $e^{(1)},...,e^{(N)}$ in the sample, and we stratify the sequence into $K$ number of bins such that the bin $s$ has $N r_s$ number of propensity scores (corresponding subjects) where vector $(r_1,...,r_K)$ satisfies $\sum_s r_s=1$. And for each subject define the variable $S \in \{1,...,K\}$ to denote its propensity score bin, i.e., $e^{i}$ is related with some $S=s$. Then the bin $s$ has many different propensity score values but in the stratification we assume that they can be represented by a single score, say, $e^s$, for $s=1,...,K$. Then by estimating unknown $e^s$ with the proportion of treated subjects in the bin $s$, i.e., $e^s=N_{1s}/N r_s$ where $N_{1s}$ and $N_{0s}$ are number of treated and untreated subjects that belong to the bin $s$, so $Nr_s =N_{1s}+N_{0s}$. Then 

 \begin{eqnarray*}
\rho & =& \frac{1}{N} \sum_i \frac{Z^i Y^i}{e^i} -\frac{1}{N} \sum_i \frac{(1-Z^i) Y^i}{1-e^i} \\
  & \approx & \frac{1}{N} \sum_s \sum_i \frac{Z^i Y^i}{e^s} I(S_i=s)- \sum_s \frac{1}{N} \sum_i \frac{(1-Z^i) Y^i}{1-e^s} I(S_i=s)\\
& = & \sum_s r_s \sum_i \frac{Z^i Y^i}{N_{1s}} I(S_i=s)- \sum_s r_s \sum_i \frac{(1-Z^i) Y^i}{N_{0s}} I(S_i=s) = \rho_s
\end{eqnarray*}
which is the stratified estimator. Due to these approximations stratified estimator may not be equal to \emph{IPTW} estimator. Usually in practice $K=5$, therefore in the estimator there are only $5$ possible values of propensity scores are used even though there should be $N$ number of  propensity scores.

\subsection{Doubly Robust Estimator}

So called doubly robust (\emph{DR}) estimator (see \cite{LD2004} and references therein) is a popular one in potential outcome framework. To understand how it is related to graphical model estimator let us suppose the case that in the causal network we use predicted outcome, say, $\hat{Y}$ instead of what is really observed $Y$; that is we can use two separate regression model, say, $\hat{Y}_1:=E\{Y \vert Z=1,X \}$ and $\hat{Y}_0:=E\{Y \vert Z=0,X \}$ to predict possible outcomes for each subject. By this task which is done external to the causal graphical model, we have data for a pair of variables $\hat{Y}_0$ and $\hat{Y}_1$ for $Z=0$ and $Z=1$ respectively for each subject even though each subject has either $Z=0$ or $Z=1$. Firstly for simplicity let us assume that both $\hat{Y}_0$ and $\hat{Y}_1$ take only values from the set $\{0,1\}$ (as if the regression functions are classifiers). Then, the average causal effect estimate based on predicted outcome, say, $\rho_p$;

\begin{eqnarray*}
\rho_p &= & \sum_y y p(\hat{Y}=y \vert do(Z=1)) - \sum_y y p(\hat{Y}=y \vert  do(Z=0)) \\
      &=& \sum_y y \sum_x p(\hat{Y}=y \vert Z=1,X=x)p(X=x) - \sum_y y \sum_x p(\hat{Y}=y \vert Z=0,X=x)p(X=x) \\
    &=& \frac{1}{N} \sum_i \frac{Z^i \hat{Y}_1^i}{e^i} -\frac{1}{N} \sum_i \frac{(1-Z^i) \hat{Y}_0^i}{1-e^i} 
\end{eqnarray*}
Note that the above estimator is dependent of the used regression models.  One drawback of the $\rho_p$ is that it is not using both predictions for each subject even though both are available. So, let us consider the following modification to it to get another estimate, say, $\rho_p'$;
\begin{eqnarray*}
\rho_p'   &=& \rho_p - \Bigg\{ \frac{1}{N}\sum_{i} \hat{Y}_1^i  - \frac{1}{N}\sum_{i} \hat{Y}_0^i  \Bigg\}    \\
            &=& \frac{1}{N} \sum_i  \Bigg\{ \frac{Z^i \hat{Y}_1^i}{e^i} - \hat{Y}_1^i \Bigg\} -\frac{1}{N} \sum_i \Bigg\{ \frac{(1-Z^i) \hat{Y}_0^i}{1-e^i} - \hat{Y}_0^i \Bigg\} \\
           &=& \frac{1}{N} \sum_i \frac{(Z^i-e^i) \hat{Y}_1^i}{e^i} +\frac{1}{N} \sum_i \frac{(Z^i-e^i) \hat{Y}_0^i}{1-e^i}
\end{eqnarray*}
Now 
\begin{eqnarray*}
\rho - \rho'_{p}&=& \frac{1}{N} \sum_i  \Bigg\{ \frac{Z^i Y^i}{e^i} - \frac{(Z^i-e^i) \hat{Y}_1^i}{e^i} \Bigg\} - \frac{1}{N} \sum_i \Bigg\{ \frac{(1-Z^i) Y^i}{1-e^i} + \frac{(Z^i-e^i) \hat{Y}_0^i}{1-e^i} \Bigg\} = DR
\end{eqnarray*}
which is called the doubly robust estimator ($DR$). That is, we can have the $DR$ estimator from the graphical model estimator if we use both the observed outcome and some predicted outcome in the graphical model. Note that $\rho_p'$ can be effectively zero if our propensity score estimates are equal to respective sample proportions i.e., the maximum likelihood estimates from the  joint likelihood for $p(y,z,x)$. Then we get the $DR$ and the $IPTW$ the same in this case. Furthermore  numerically the $IPTW$ is just the maximum likelihood parameter estimate based graphical model estimate when the propensity scores are sample proportions. So we have that the $DR$ is numerically equal to the basic graphical model estimator in this case. 

Often researchers estimate propensity scores through a model, for example, a logistic regression with independent variables $X$ (as a linear or/and non-linear combination of them). But generally no one knows the true model in a given empirical context therefore $DR$ estimate may be affected by the propensity model specification. When propensity scores are consistently estimated the  $IPTW$ estimator is a consistent to the average causal effect, therefore so does the $DR$  estimator. Note that the result is true irrespective of  specification of the two regression models $E\{Y \vert Z=1,X\}$ and $E\{Y \vert Z=0,X\}$ -whether they are true or not. However for small samples, $DR$ estimate may depend on the used regression models if estimated  propensity scores are different from corresponding sample proportions.

Likewise it may be of interest to see that what can the $DR$ estimator be if we have true outcome regression models. First  consider the following writing of the $DR$ estimator.
\begin{eqnarray*}
DR &=& \frac{1}{N} \sum_i  \Bigg\{ (Y^i - \hat{Y}_1^i )   \frac{Z^i }{e^i} -  (Y^i - \hat{Y}_0^i ) \frac{(1-Z^i)}{1-e^i}  \Bigg\} +  \Bigg\{ \frac{1}{N}\sum_{i} \hat{Y}_1^i -\frac{1}{N}\sum_{i} \hat{Y}_0^i  \Bigg\}
\end{eqnarray*}
And from the above we know that  $E_x E_y \{ Y \vert Z=1,X\} = \sum_{x,y} yp(y \vert Z=1,x)p(x)= \frac{1}{N} \sum_i  \frac{Y^i Z^i}{e^i} $ and therefore  $E_x E_{\hat{y}} \{ \hat{Y} \vert Z=1,X\}=\sum_{x,\hat{y}} \hat{y}p(\hat{y} \vert Z=1,x)p(x)= \frac{1}{N} \sum_i  \frac{\hat{Y}^i Z^i}{e^i} $.  Now consider the case of $Z=1$. Since for each $X=x$, $\hat{Y}$ is a single value, say, $\hat{y}(Z=1,x)$ then we have $\sum_{x,\hat{y}} \hat{y}p(\hat{y} \vert Z=1,x)p(x) = \sum_x \hat{y}(Z=1,x) p(x)$. If we take $\hat{y}(Z=1,x)=\sum_{y} yp(y \vert Z=1,x)$ for each $x$, i.e., if we let our regression function at $X=x$ to be the empirical mean of $Y$ values at $X=x$ then we get $\sum_i  \frac{Y^i Z^i}{e^i}=\sum_i  \frac{\hat{Y}^i Z^i}{e^i}$. And in the similar way, for the case of $Z=0$ we get $\sum_i  \frac{Y^i (1-Z^i)}{1-e^i}=\sum_i  \frac{\hat{Y}^i (1-Z^i)}{1-e^i}$. Both of them imply that
\begin{eqnarray*}
DR &=&   \frac{1}{N}\sum_{i} \hat{Y}_1^i -\frac{1}{N}\sum_{i} \hat{Y}_0^i  \\
     &=&  \frac{1}{N}\sum_{x} N_x \hat{Y}(Z=1,x) - \frac{1}{N}\sum_{x} N_x \hat{Y}(Z=0,x) \\
      &=&  \sum_{x} \hat{Y}(Z=1,x) p(x)- \sum_{x}  \hat{Y}(Z=0,x) p(x)
\end{eqnarray*}
That is, if the outcome regression model $\hat{Y}(Z=z,X=x)$ has its value at $X=x$ as the mean of the observed $Y$ values at $X=x$ for $Z=z$, for $z=0,1$ or in other words when our two regression models are the true models then the $DR$ estimator has the above simple form, that is independent of propesity score model -whether it is correct or not.

In fact we do not need to have above restriction on values of $\hat{Y}_0$ and $\hat{Y}_1$ for the validity of the above discussion. Generally a regression model predicts a continuous variable and for any continuous random variable $Y$ when we have a random sample of $n$ observations, say $\{ y_1,...,y_n \}$, $ \int_y y p(y)$ is estimated by $\sum_{i=1}^n y_i /n$. In general when $Y$ is continuous and $X$ is mixture of discrete and  continuous then writing all summations of $X$ as integration, if any,  we have that 
\begin{eqnarray*}
\int_x \int_y yp(y \vert z,x)p(x) dydx&=& \int_x \frac{1}{N(z,x)} \sum_j y^jI(Z=z)I(X=x) \\
                                             &=& \frac{1}{N} \sum_x \frac{N(x)}{N(z,x)} \sum_j y^jI(Z^j=z)I(X^j=x)  \\
                                             &=& \frac{1}{N} \sum_j   y^j   \sum_x \frac{I(Z^j=z)I(X^j=x)}{N(z,x)/N(x)}  \\
                                            &=& \frac{1}{N} \sum_j   y^j   \frac{I(Z^j=z)I(X^j=x^j)}{P(Z^j=z \vert X^j=x^j)}
\end{eqnarray*}
Therefore $\int_x \int_y yp(y \vert Z=1,x)p(x)=\frac{1}{N} \sum_j   \frac{y^j z^i}{e^j}$ and $ \int_x \int_y yp(y \vert Z=0,x)p(x)=\frac{1}{N} \sum_j     \frac{y^j (1-z^i)}{1-e^j}$. So, when $\hat{Y}_0$ and $\hat{Y}_1$ are continuous random variables then $\rho_p = \int_y y P(\hat{Y}=y \vert do(Z=1)) - \int_y y P(\hat{Y}=y \vert  do(Z=0))$ that can be estimated with summations, thus above formulas can be obtained. From above it is clear all the discussion can be generalized to the case of when $Y$ and $X$ have any finite state spaces.

{}

\end{document}